# A thin anisotropic metasurface for simultaneous light focusing and polarization manipulation

Mehdi Veysi, Ozdal Boyraz, and Filippo Capolino

*Department of Electrical Engineering and Computer Science, University of California Irvine, Irvine, California 92697, USA*
*f.capolino@uci.edu*

*Abstract*- The possibility of integration of two important categories of optical components, i.e., circular polarizer and lens, into a thin plasmonic metasurface is examined. After exploring general theoretical formulation, optimal designs to realize polarizing lenses at mid-infrared frequencies are presented, for the cases when metal losses cannot be neglected. A two-dimensional array of Y-shaped nanoantennas with polarization dependent and spatially varying phase response is the basis for such a unique compact polarizing lens, whose properties can have a high impact in integrated optics.

CONVENTIONAL optical devices such as lenses with aberration correction, and quarter-wave plate made of birefringent and chiral materials met the performance demand in both bandwidth and efficiency, but they are usually bulky, expensive and difficult to fabricate and to integrate in nanophotonic systems. Nevertheless, applications demand cheaper and thinner devices with better performance. In the last few years, subwavelength plasmonic metasurfaces (MSs) have attracted increasing attention in modern optics and photonics due to their capability to introduce phase change discontinuities over subwavelength distances [1-4]. Techniques of beam shaping through phase discontinuities achieved by collections of resonant antennas have been used at centimeter and millimeter waves for a few decades [5]. This idea, applied to visible/ infrared frequencies, can eliminate the need for propagation path compensation in lensing [6-9], or reduce the physical dimensions required for a quarter-wave plate [10-12]. Since the circular polarizer and lens are at the heart of many optical systems, the possibility of having both light focusing and polarization manipulation capabilities with a thin single layer plasmonic MS would significantly reduce the cost, volume, optical loss, and system complexity.

Recently, single [13] and multilayer [14] MSs, made of Au layers with negligible losses, have been introduced at mid-infrared frequencies for both phase and polarization control. However, in various cases, especially at infrared and optical frequencies, materials have unavoidable losses, which need to be carefully addressed in order to ensure a practical design. In general, in mid-infrared and far-infrared regime, the variations in the amplitude of the reflection/transmission coefficient with respect to the nanoantenna design parameters are negligible when low loss metals, such as Au and Ag, are used. In this case, one needs to account only for the phase of the reflection/transmission coefficient. However, since Au and Ag are lossy at visible and near-infrared regime, the amplitude of the reflection/transmission coefficient significantly changes with the nanoantenna design parameters. So, the synthesis of the polarizing lens at visible/infrared regime is not in general a phase only synthesis method and one should carefully consider both the amplitude and phase of the reflected/transmitted wave at the MS.

In this paper, we develop a phase-amplitude synthesis method for the polarizing lens design. We show a novel single layer metalens design capable of focusing and polarization manipulation, simultaneously. The theory and the examples developed here will enable improvement of several devices, such as (i) reflecting or (ii) transmitting focusing lenses, (iii) polarizing lenses, (iv) lenses with dual foci, one for each polarization, and (v) lenses with dual foci one for each wavelength. The first challenge is to find an appropriate nanoantenna, which can simultaneously satisfy both a wide reflection phase range and polarization conversion. Hence, an anisotropic nanoantenna, which allows for independent tuning of the phase changes experienced by the $x$- and $y$-pol incident fields, is needed. The nanoantennas are made of low cost aluminum with non-negligible loss at mid-infrared to account for the loss in the design process. Although, we focus on the reflection-type geometry, the theory can be extended to the transmission-type geometry.

The proposed polarizing lens design consists of an array of aluminum (Al) Y-shaped nanoantennas patterned on one side of a silica substrate with thickness of 400nm deposited onto aluminum ground plane as illustrated in Fig. 1a. The MS is in the $x$-$y$ plane, and it is illuminated by a normally incident slant-polarized plane wave whose transverse–to-$z$ electric field phasor is $\mathbf{E}_t^{inc} = E_0\,(\hat{\mathbf{x}}\cos\varphi_i + \hat{\mathbf{y}}\sin\varphi_i)\,e^{jkz}$, where $E_0$ is its magnitude, $k$ is the free-space wavenumber, and $\varphi_i$ is the angle of the electric field vector with respect to the *x*-axis (a $e^{j\omega t}$ time harmonic convention is assumed). The electric field of the wave reflected by the polarizing lens is easily decomposed into two circularly polarized fields, whose transverse components are

$$\mathbf{E}_t^r = \frac{1}{\sqrt{2}} E_0 |\Gamma_x| \cos\varphi_i \, e^{-jkz} e^{j\phi_x} [\hat{\mathbf{e}}_R(1+C) + \hat{\mathbf{e}}_L(1-C)] \quad (1)$$

$$\hat{\mathbf{e}}_L = \frac{(\hat{\mathbf{x}}+j\hat{\mathbf{y}})}{\sqrt{2}}, \quad \hat{\mathbf{e}}_R = \frac{(\hat{\mathbf{x}}-j\hat{\mathbf{y}})}{\sqrt{2}}, \quad C = \left|\frac{\Gamma_y}{\Gamma_x}\right| \tan\varphi_i \, e^{j(\phi_y-\phi_x+\pi/2)}$$

where $\Gamma_{x,y} = |\Gamma_{x,y}| e^{j\phi_{x,y}}$ is the reflection coefficient pertaining to the *x*- and *y*-pol components, respectively. In order to change the polarization state of the incident wave, a plasmonic MS made of anisotropic Y-shaped elements is employed. At a certain wavelength in which the phase difference between *y*- and *x*-pol reflection coefficients, $(\phi_y - \phi_x)$, is 90° and the ratio of the wave reflection magnitude for the *y*-pol to that of the *x*-pol is almost constant, $|\Gamma_y|/|\Gamma_x| = \cot\varphi_i$, the reflected wave is purely left-hand circularly polarized (LHCP). It is purely right-hand circularly polarized (RHCP) when $(\phi_y - \phi_x) = -90°$.

In order for a MS to generate focusing, the reflected field's phase distribution needs to be varied with the *x, y* position. For this purpose, let us denote the MS element locations with $\mathbf{r}^{mn} = \mathbf{r}^{00} + ma\hat{\mathbf{x}} + nb\hat{\mathbf{y}}$ where *m* and *n* are integers, *a* and *b* denote the local periods along *x* and *y*. Under plane wave incidence propagating in the direction $\hat{\mathbf{k}}$, the required phase of the reflection coefficient at a $mn^{\text{th}}$ MS unit cell, $\phi^{mn}$, must satisfy the generalized Snell's law [5]

$$\phi^{mn} - k_0 \left[|\mathbf{r}^{mn} - \mathbf{r}_f| + \mathbf{r}^{mn} \cdot \hat{\mathbf{k}}\right] = 2p\pi + c, \quad p = 0,1,.. \quad (2)$$

in order to focus the reflected wave at $\mathbf{r}_f$. Here, *c* is an arbitrary constant phase, and the subscripts *x* and *y* have been dropped since this equation applies to both polarizations. In summary, the focusing and polarization conversion mechanism of the anisotropic plasmonic metalens under the linear polarization illumination is conceptually described by Eq. (1)-(2), where two approximations extensively used in reflectarray research have been applied [5, 8, 15]: (i) the concept of local periodicity, and (ii) the reflected phases $\phi_{x,y}^{mn}$ are evaluated at normal incidence, for each unit cell. The latter approximation has been considered accurate especially for small to moderate incidence angles and focal length to diameter ratio larger than unity (in our case *f*/D=1.25). Note that all possible polarizations (linearly polarized (LP), RHCP, and LHCP) can be achieved by properly adjusting the $\varphi_i$ angle. To independently vary the reflection phases $\phi_{x,y}$ and amplitudes $|\Gamma_{x,y}|$ of each principle polarization of the incident wave, along the *x* and *y* directions, each unit cell should possess enough degrees of freedom. Hence, a Y-shaped nanoantenna shown in Fig. 1, recently used for thermal tuning the resonances of mid-infrared (IR) plasmonic antennas [16], is utilized as a unit cell. Its symmetric and asymmetric current modes are depicted in Fig. 1(a). Varying $\Delta$ and $\ell_1$ leads to changes in both the *x*- and *y*-directed current paths. Note that $\ell_2$ affects only the extension of the *y*-directed surface current path and has a negligible effect on the asymmetric mode.

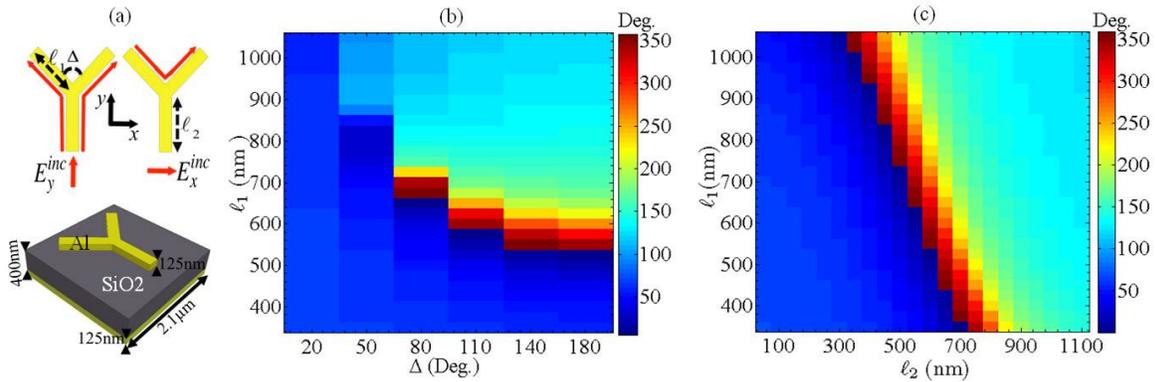

Fig. 1. (a) Schematic of a MS Y-shaped nanoantenna together with its current distribution for symmetric (top left) and asymmetric modes (top right), and the 3D view of each MS cell (bottom). (b) The reflection phase of the asymmetric mode for a MS made of identical Y-shaped nanoantennas at 4.3μm as a function of the arm length, and the angle between the two arms, $\Delta$, for $\ell_2 = 50$nm. (c) The reflection phase of symmetric mode for a MS made of identical Y-shaped nanoantennas at 4.3μm as a function of the arm length, and stub length, for $\Delta=110°$. The width of each arm is fixed at 200nm, and the complex relative permittivities of the Al and SiO2 at 4.3μm are -1601.3+j609.4 [17] and 1.5, respectively.

In the design, first, the two physical parameters of the Y-shaped elements, arm length $\ell_1$ and arm angle $\Delta$, are tuned to satisfy the focusing condition in Eq. 2 for the x-pol incident wave at each $mn^{th}$ location on the MS as

$$err_{(1)}^{mn}(\Delta, \ell_1) = \left| \phi_x^{mn} - \phi_x^{simulated}(\Delta, \ell_1) \right| \quad (3)$$

here $\phi_x^{simulated}$ is phase of the reflected x-pol at the design wavelength λ, obtained from a simulation with the geometrical parameters optimized to meet the required $\phi_x^{mn}$ from Eq. 2. The elements are optimized using the frequency domain finite-element method (FEM) (provided by HFSS by Ansys Inc.) in a fully periodic arrangement (based on local periodicity assumption). To achieve circularly polarized focused beam upon reflection, it is necessary to tailor both the spatial phase and amplitude distributions of the reflected field by tuning each element on the MS. Therefore, the stub length $\ell_2$ is then tuned to meet the proper condition for converting the polarization from linear to circular and to meet the focusing condition in Eq. 2 also for the y-pol. Note that tuning the stub length $\ell_2$ has a negligible effect on the asymmetric mode of the Y-shaped elements and thus on the x-pol focusing. Therefore, a second error function is defined based on the axial ratio (AR) of the reflected wave as

$$err_{(2)}^{mn}(\Delta, \ell_1, \ell_2) = \left| \pm 1 - AR(\Delta, \ell_1, \ell_2) \right| \quad (4)$$

where plus or minus signs stands for a desired LHCP or RHCP, respectively. The AR depends on the physical parameters of the elements and is expressed as

$$AR(\Delta, \ell_1, \ell_2) = -\frac{|E_R(\Delta, \ell_1, \ell_2)| + |E_L(\Delta, \ell_1, \ell_2)|}{|E_R(\Delta, \ell_1, \ell_2)| - |E_L(\Delta, \ell_1, \ell_2)|} \quad (5)$$

where $E_R$ and $E_L$ are projections of the simulated $\mathbf{E}_t^r$, (see Eq. 1) onto the RHCP and LHCP unit vectors $\hat{\mathbf{e}}_R$ and $\hat{\mathbf{e}}_L$, respectively. Note that for each element the phase difference $\left| \phi_x^{mn} - \phi_y^{mn} \right|$ between the x- and y-pol reflection coefficients is set very close to 90° when minimizing $err_{(2)}^{mn}$ in Eq. (4). Therefore, the focusing condition for $\phi_y^{mn}$ in Eq. 2 is automatically satisfied provided that $\phi_x^{mn}$ satisfies it, as already imposed by Eq. (3). The range of phases of the x- and y-pol reflection coefficients for a MS made of identical Y-shaped nanoantennas as a function of the physical parameters (arm length, stub length, and arm angle) at λ= 4.3µm are shown in Fig 1(b-c). Based on these results, and using the theory described in this paper, a square, flat polarizing lens with dimensions 28µm×28µm (6.51λ×6.51λ) with focal length $f$=35µm (numerical aperture (NA)=0.37), and $α$=30° is designed. The beam waist of the incident Gaussian wave ($w_0$=24µm) is chosen so that almost the entire surface of the polarizing lens is illuminated by a plane-wave-like wavefront. Fig. 2 shows the schematic of the simulation setup together with the intensity of the scattered field in the longitudinal and transverse planes, calculated by FEM full wave simulations. The focus is clearly observed. The simulation results for the axial ratio of the scattered wave along the y transverse direction, in the focal plane, is also plotted in Fig. 2(c). The axial ratio is around 3dB in the focal region. Increasing the size of the MS not only reduces the effect of undesirable edge diffraction and specular reflection on both focusing and polarization conversion but also increases the number of the nanoantennas for which the local periodicity assumption used in the design process is accurate [5]. This would improve the lens performance in terms of both focusing and polarization conversion. Thus, a square, flat polarizing lens with dimensions 100µm×100µm (23.25λ×23.25λ), focal length $f$=125µm (NA=0.37) and $α$=30⁰ is also designed. Due to excessive full wave simulation requirements, and based on the results in Fig. 1, the field intensity and AR in the focal region generated by this metalens is analytically computed using the reciprocity theorem and the Fourier transform of the aperture field [18].

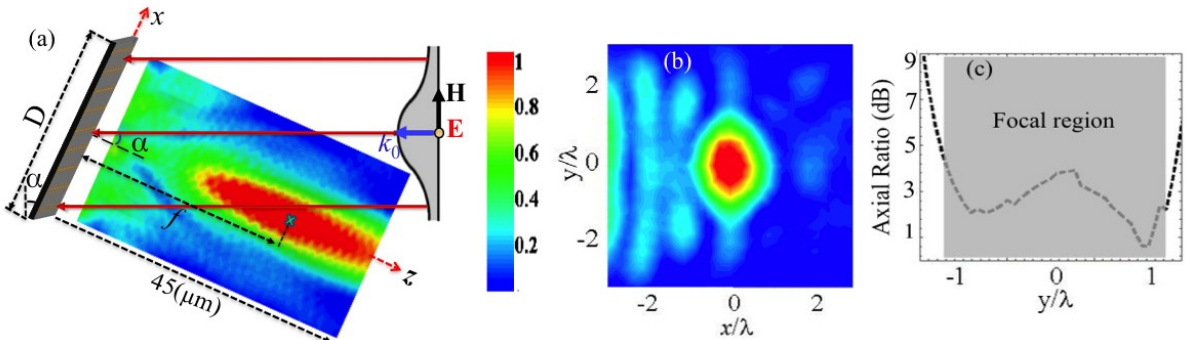

Fig. 2. (a) Simulation setup together with the full wave simulated results of the normalized scattered field intensity ($|\mathbf{E}|^2/|\mathbf{E}_0|^2$) in x-z plane at λ=4.3µm: for a 28µm by 28µm sized flat polarizing lens with $f$=35µm, and $α$=30°, (b) the normalized intensity of the scattered field in x-y transverse focal plane. (c) Axial ratio of the scattered wave along a horizontal line in the focal plane (x=0, z=35µm), the highlighted rectangle shows the focus region.

Note that, in contrast to the full wave method that precisely account for the edge diffraction and specular reflection effects, the Fourier transform method does not. The bigger the size of the MS, the better the agreement between full wave and Fourier transform methods. The electric field intensity and axial ratio in the focal plane are shown in Fig. 3 that show an improvement compared to Fig. 2. Now the focus beam has a spot size of less than 7.7µm, in the *x-y* plane, very close to the Abbe's diffraction limit, 6.97µm (0.6λ/NA), with an axial ratio close to unity. The designed MS focuses nearly all the incident energy arriving from the 30° elevation angle onto a small spot centered at the focal point. However, as the incident beam angle moves away from 30°, the focus spot starts broadening out and moving away from the desired focal point.

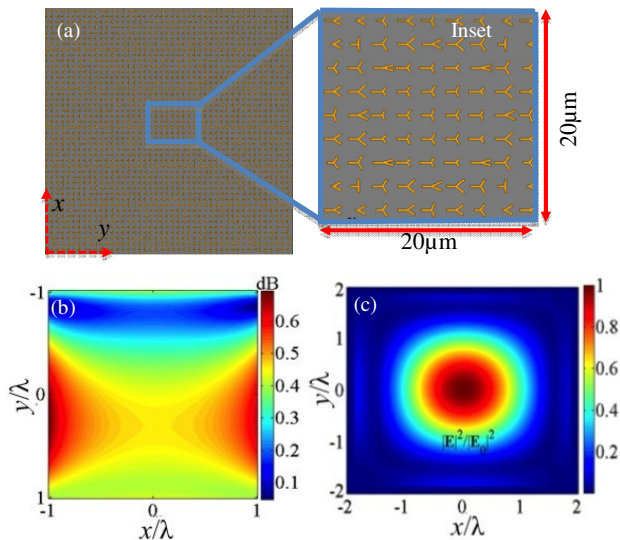

Fig. 3. (a) Schematic of a large-size polarizing lens (23λ×23λ) at λ=4.3µm with: *D*=100µm, *f*=125µm, *α*=30˚, *θ_i*=53˚, together with its simulation results (b) normalized field intensity at the focal plane, and (c) AR in the focused region.

In conclusion, the concept of flat polarizing lens, capable of simultaneous polarization and focusing manipulation, based on a flat MS is demonstrated. A planar array of anisotropic Y-shaped nanoantennas is employed to generate light focusing and polarization state conversion. The integration of polarizer and lenses into a single layer thin MS would significantly reduce cost, volume, optical loss, and system complexity.

ACKNOWLEDGMENT

The authors would like to thank Caner Guclu (University of California, Irvine) for fruitful discussions. We are also grateful to ANSYS, Inc., for providing HFSS.